\documentclass[twocolumn,secnumarabic,amssymb, nobibnotes]{revtex4-2}

\usepackage{graphicx}% Include figure files
\usepackage{dcolumn}% Align table columns on decimal point
\usepackage{bm}
\usepackage[colorlinks=true,allcolors=blue]{hyperref}
\usepackage{float}
\usepackage{physics}
\usepackage{slashed}
\usepackage{relsize}
\usepackage{amsmath}
\usepackage{hyphenat}

\begin{document}

\title{Proton Structure from a Soft-Wall Holographic QCD Model: Mass Spectrum, Form Factors, and Mechanical Properties}%

\author{Jiali Deng}%
\email{djl2022010355@mails.ccnu.edu.cn, houdf@mail.ccnu.edu.cn}
\affiliation{Institute of Particle Physics and Key Laboratory of Quark and Lepton Physics (MOS),
 Central China Normal University, Wuhan 430079, China}
 \author{Defu Hou}
\affiliation{Institute of Particle Physics and Key Laboratory of Quark and Lepton Physics (MOS),
 Central China Normal University, Wuhan 430079, China}
\date{\today}%
\begin{abstract}
 
 Understanding the internal structure of the proton—including its mass spectrum, electromagnetic and gravitational form factors, and mechanical properties—remains a central challenge in hadronic physics. While lattice QCD and experimental measurements provide valuable insights, a holographic framework with a single parameter set capable of simultaneously describing these diverse observables is still lacking. Here, we employ the soft wall model, a phenomenological holographic approach that incorporates gluon condensation and linear confinement, to compute the proton mass spectrum, electromagnetic form factors (EMFFs), and gravitational form factors (GFFs). Our results show good agreement with recent experimental data and lattice QCD calculations. Despite its phenomenological nature, the model's ability to simultaneously describe multiple observables suggests that it effectively mimics some key QCD features. 

\end{abstract}
\maketitle
%\tableofcontents

\section{Introduction}\label{sec:01_intro}

What constitutes the mass and spatial extent of a proton? This deceptively simple question lies at the heart of one of the most profound challenges in modern physics: understanding the non-perturbative structure of hadrons governed by Quantum Chromodynamics (QCD). As the primary building block of visible matter, the proton's internal dynamics—dictated by the complex interplay of confined quarks and gluons—elude a complete analytical description from first principles. Resolving this puzzle is not merely an academic pursuit but is fundamental to advancing our comprehension of nuclear physics, astrophysical phenomena, and the Standard Model itself. A comprehensive mapping of the proton's properties, including its mass spectrum, electromagnetic and gravitational form factors (GFFs), and charge radii, provides critical windows into this sub-femtoscale domain. These observables encode information on the spatial distribution of energy, momentum, and stress within the proton, thereby elucidating the mechanics that govern its interactions and stability \cite{Guidal:2013rya, Selyugin:2014sca, Burkert:2023wzr}.

Traditional lattice QCD simulations have made significant strides in computing static hadronic properties from the discretized QCD Lagrangian. However, extracting real-time dynamical processes and form factors at low momentum transfer remains computationally intensive and often limited by systematic uncertainties. Similarly, while perturbative QCD is highly successful in describing high-energy scattering, it fails in the non-perturbative regime where the coupling constant becomes large. This gap in our calculational toolkit has spurred the development of alternative non-perturbative frameworks.  In this non-perturbative frontier, holographic QCD has emerged as a powerful complementary framework. Rooted in the anti-de Sitter/Conformal Field Theory (AdS/CFT) correspondence, it posits a duality between a strongly coupled gauge theory in four dimensions and a classical gravitational theory in a higher-dimensional curved spacetime \cite{Maldacena:1997re, Witten:1998qj,Aharony:1999ti}. Although the original correspondence was formulated for a maximally supersymmetric, conformal gauge theory, its conceptual and mathematical toolkit has been successfully adapted, through various phenomenological models, to approximate key features of real-world QCD, such as confinement and chiral symmetry breaking \cite{Polchinski:2001tt, Karch:2006pv,Deng:2021kyd,He:2007juu}. These "bottom-up" holographic models, particularly the soft-wall model, provide a tractable geometric language to calculate hadronic properties that are notoriously difficult to access via lattice QCD or other first-principles methods, making them invaluable for generating testable predictions and conceptual insights.

Significant progress has been made using these approaches. The proton's mass spectrum, a foundational benchmark, has been computed within various holographic and other theoretical frameworks, establishing a baseline for model calibration \cite{deTeramond:2005su, Brodsky:2014yha, FolcoCapossoli:2019imm, Deng:2025fpq}. Similarly, electromagnetic form factors—the Dirac ($F_{1}$) and Pauli ($F_{2}$) form factors—which encode the spatial distributions of the proton's charge and magnetization \cite{Pacetti:2014jai, Punjabi:2015bba}, have been extensively studied both experimentally via elastic electron-proton scattering \cite{Arrington:2007ux, Puckett:2011xg, Xiong:2023zih} and theoretically within holography \cite{Abidin:2009hr, Sufian:2016hwn, Mamo:2021jhj, Ahmady:2021qed}. These form factors provide critical, but incomplete, information about the proton's static structure.

A more comprehensive picture requires the gravitational form factors (GFFs), which describe the proton's coupling to the energy-momentum tensor (EMT). These form factors are of paramount importance as they reveal the origins of the proton's mass and spin, and map its internal distributions of energy, pressure, and shear forces \cite{Brodsky:2000ii, Burkert:2023wzr}. They address the so-called "proton mass and spin crises" by quantifying how these quantum numbers are partitioned among quarks and gluons. While analogous in concept to electromagnetic form factors—probing gravitational rather than electromagnetic interactions—GFFs are far more challenging to access experimentally. Pioneering calculations for mesons and the nucleon have been performed in holographic QCD \cite{Abidin:2008ku, Mamo:2021krl, Hackett:2023rif, Nair:2024fit, Dehghan:2025ncw}, providing theoretical benchmarks and highlighting the rich mechanical structure hidden within the proton.

Despite these advances, a bottom-up soft-wall holographic QCD model that achieves a globally accurate and statistically improved description of the proton mass spectrum, electromagnetic form factors, and gravitational form factors simultaneously—compared to existing soft-wall implementations \cite{Mamo:2021jhj,Mamo:2021krl}—has not yet been fully explored. Such an integrated analysis provides a useful check on the descriptive capacity of a given holographic model, by testing whether a single set of calibrated parameters can describe a range of proton properties. In this work, we provide such a description within the soft-wall framework by introducing a modified dilaton field.

In this work, we employ a phenomenological holographic model with a single parameter set, which incorporates key QCD features like linear Regge trajectories and gluon condensation, to analyze the proton's structure. Our specific objectives are: (1) to compute the proton's mass spectrum within this model and determine its fundamental parameters; (2) to calculate the proton's electromagnetic form factors and the corresponding charge/magnetic radii; (3) to derive the gravitational form factors and mechanical properties. 

The remainder of this paper is structured as follows. Section II introduces the soft-wall holographic QCD framework and details the calculation of the proton mass spectrum. Section III presents the computation of the proton's electromagnetic form factors. The gravitational form factors are derived and analyzed in Section IV. Finally, Section V summarizes our key findings and discusses their broader implications.

\section{Mass spectrum for proton}\label{sec:02}

The metric of five-dimensional AdS spacetime is given by
\begin{equation}
	\label{eq1}
ds^{2}=g_{mn}dx^{m}dx^{n}=\frac{1}{z^2}(\eta_{\mu\nu}dx^{\mu}dx^{\nu}-dz^2),
\end{equation}
where $\eta_{\mu\nu}=diag (1,-1,-1,-1)$ and $\mu, \nu=0,1,2,3$. The fifth-dimensional coordinate z extends from $z\rightarrow 0$ (the ultraviolet/UV boundary) to $z\rightarrow \infty$ (the infrared/IR boundary).

In the holographic framework, the proton is described by a dual massive spinor field propagating in five-dimensional anti-de Sitter (AdS₅) space. The corresponding action for this spinor field is given by \cite{Abidin:2009hr}
\begin{equation}
\begin{split}
	\label{eq2}
S=&\int d^{4}xdz\sqrt{g}e^{-\Phi(z)}(\frac{i}{2}\overline{\Psi}e^{N}_{A}\Gamma^{A}D_{N}\Psi\\
& -\frac{i}{2}(D_{N}\Psi)^{\dag}\Gamma^{0}e^{N}_{A}\Gamma^{A}\Psi-(m_5+\Phi(z))\overline{\Psi}\Psi),
\end{split}
\end{equation}
where $e^{N}_{A}=z\delta^{N}_{A}$ represents the inverse vielbein, $D_{N}=\partial_{N}+\frac{1}{8}\omega^{ab}_{N}[\Gamma_{a},\Gamma_{b}]-iV_N$ denotes the covariant derivative and $m_5$ represents the five dimensional mass of the proton. The Dirac gamma matrices $\Gamma^{A}=(\gamma^\mu,-i\gamma^5)$ satisfy the Clifford algebra: $\gamma^{a}\gamma^{b}+\gamma^{b}\gamma^{a}=2\eta^{ab}$. $\omega^{ab}_{N}$ represents the spin connection:
\begin{equation}
	\label{eq3}
\omega^{ab}_{N} = e^{a}_{M} \partial_N e^{bM} + e^{a}_{M} \Gamma^{M}_{NP} e^{bP}.
\end{equation}

We take the dilaton field in the form of \cite{Li:2013oda}
\begin{equation}
	\label{eq3}
\Phi(z)=k_1^2z^2tanh(k_2^4z^2/k_1^2).
\end{equation}
In this formulation, the dilaton field exhibits the following behavior at the UV boundary:
\begin{equation}
	\label{eq4}
\Phi(z)=k_2^4z^4,
\end{equation}
and corresponds holographically to the dimension-4 gauge-invariant gluon condensate in the boundary field theory, while at IR boundary takes the following form
\begin{equation}
	\label{eq5}
\Phi(z)=k_1^2z^2.
\end{equation}
This form gives rise to linear confinement.

The equation of motion for the spinor field is given by
\begin{equation}
	\label{eq6}
[ie^{N}_{A}\Gamma^{A}D_{N}-\frac{i}{2}\partial_{N}\Phi(z)e^{N}_{A}\Gamma^{A}-(m_5+\Phi(z))]\Psi=0.
\end{equation}

The spinor field is decomposed into left-handed and right-handed components:
\begin{equation}
	\label{eq7}
\Psi(x^\mu,z)=(\frac{1+\gamma^5}{2}\chi_{R}(z)+\frac{1-\gamma^5}{2}\chi_{L}(z))\Psi_{(4)}(x^\mu),
\end{equation}
where $\Psi_{(4)}(x^\mu)$ satisfies the four-dimensional Dirac equation. By setting $\chi_{R/L}(z) = e^{-2A(z)+\Phi(z)/2}\varphi_{R/L}(z)$ and combining with the equation of motion, we obtain
\begin{equation}
\begin{split}
	\label{eq8}
-\varphi_{R/L}''(z)+&[(m_{5}+\Phi(z))^{2}e^{2A(z)}\pm ((m_{5}+\Phi(z))A'(z)\\
                  &+\Phi'(z))e^{A(z)}]\varphi_{R/L}(z)=M_{n}^{2} \varphi_{R/L}(z),
\end{split}
\end{equation}
where $M_n$ represents the mass spectrum of the proton in four-dimensional spacetime, with the quantum number $n=1$ denoting to the ground state and $n=2,3,\....$ corresponding the excited states. The wave functions satisfy the renormalization condition:
\begin{equation}
	\label{eq7}
\int \varphi^{(n)}_{R}(z)\varphi^{(m)}_{R}(z)dz=\int \varphi^{(n)}_{L}(z)\varphi^{(m)}_{L}(z)dz=\delta_{mn}.
\end{equation}

We obtain the mass spectrum of the proton by numerically solving the equation with Dirichlet boundary conditions, with the three parameters $k_1=0.450\ \mathrm{GeV}$, $k_2=0.318\ \mathrm{GeV}$, and $m_5=1.3$, which are determined by fitting the mass spectrum. In the standard AdS/CFT duality, the five-dimensional mass of the proton is $m_5=5/2$ as a combination of three free quarks. However, QCD is a non-conformal theory of strong interactions, so the five-dimensional mass needs to be corrected by incorporating the anomalous dimension $\gamma$ \cite{FolcoCapossoli:2020pks}. The five-dimensional mass used here is the effective value that takes into account the interactions.

\begin{table}[H]
	\centering
	\includegraphics[width=8.5cm]{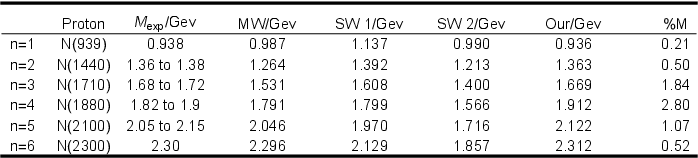}
	\caption{\label{Table 1}Comparison of the proton mass spectrum among experimental values ($M_{\text{exp}}$)\cite{ParticleDataGroup:2024cfk}, the modified warp factor model (MW)\cite{FolcoCapossoli:2019imm}, the soft-wall models (SW1 and SW2)\cite{Mamo:2021jhj, Mamo:2021krl}, and our result (Our). The deviation of our result from the central experimental value is within $3\%$ (M\%).}
\end{table}

As demonstrated in Table I, our calculated values show good agreement with experimental measurements, generally maintaining a deviation within 3\%. the maximum relative deviations of the calculated masses from the experimental values for the MW model, SW1 model, SW2 model, and our model are approximately $9.9\%$, $21.2\%$, $19.3\%$, and $2.8\%$, respectively. Compared with the traditional soft-wall model, our model modifies the behavior of the dilaton field in the intermediate-to-high energy region, which makes our model more successful in describing the proton mass spectrum. Next, using the same set of parameters, we will predict the proton's electromagnetic and gravitational form factors and compare them with these models, thereby further testing the reliability of our model.

\section{Electromagnetic form factors}\label{sec:03}

In elastic scattering processes (depicted in Fig. 1), electrons and protons interact through the exchange of a virtual photon characterized by the four-momentum transfer squared 
$q^2$. The virtual photon in this case interacts with the nucleon's constituent partons while maintaining the nucleon's integrity. Consequently, the final state preserves the nucleon identity, with momentum transfer being the sole observable change.
\begin{figure}[H]
	\centering
	\includegraphics[width=7cm]{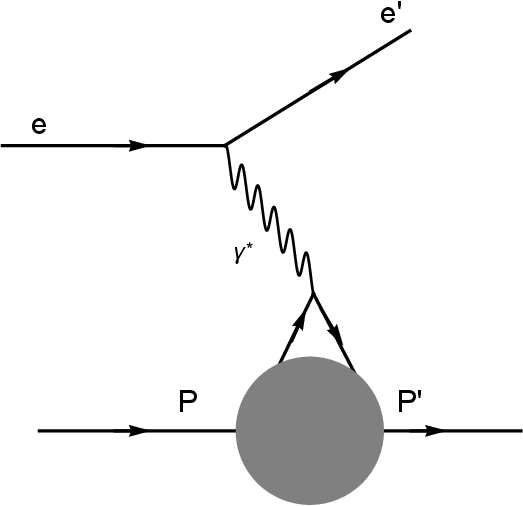}
	\caption{\label{figure}Elastic electron-proton scattering mediated by virtual photon exchange \cite{Deng:2025fpq}.}
\end{figure}

The electromagnetic form factors of protons and neutrons represent fundamental nucleon observables, playing a pivotal role in elucidating nucleon structure and dynamics. The elastic electron-proton scattering process is generally characterized by two independent form factors: the Dirac form factor $F_1$ and the Pauli form factor $F_2$
\begin{equation}
	\label{eq10}
\langle p'|J^{\mu}(0)|p\rangle=\bar{u}(p')[\gamma^{\mu}F_{1}(Q^{2})+\frac{i\sigma^{\mu\nu}q_{\nu}}{2M}F_{2}(Q^{2})]u(p),
\end{equation}
where $\sigma^{\mu\nu}=[\gamma^{\mu},\gamma^{\nu}]$, $q_{\nu}=p'_{\nu}-p_{\nu}$ is the transfer of momentum (with $q^2=-Q^2, Q^2>0$), $M$ denotes the proton mass and $J^{\mu}(0)$ represents the quark electromagnetic current. The Dirac form factor $F_1(Q^2)$ describes matrix elements that conserve nucleon spin, whereas the Pauli form factor $F_2(Q^2)$ mediates transitions involving spin flip. These form factors respectively characterize the spatial distributions of the nucleon's electric charge and magnetic moment.

The form of the interaction action is
\begin{equation}
\begin{split}
	\label{eq11}
&\int d^{4}xdz\sqrt{-g}e^{-\Phi(z)}[\eta_1\bar{\Psi}_{p'}(x,z)e^{M}_{A}\Gamma^{A}\phi_{M}(x,z)\Psi_{p}(x,z)\\
&+\eta_2\bar{\Psi}_{p'}(x,z)e^{M}_{A}e^{N}_{B}[\Gamma^{A},\Gamma^{B}]F_{MN}(x,z)\Psi_{p}(x,z)].
\end{split}
\end{equation}
where $\eta_1=0.5$ and $\eta_2=0.667$ are fixed by the constraints of charge conservation, $F_1(0)=1$, and the experimentally measured proton magnetic moment $F_2(0)=1.793$ respectively. The Dirac form factor is given by 
\begin{equation}
\begin{split}
	\label{eq12}
F_{1}(Q^{2})&=\frac{1}{2}\int dzV(Q,z)(\varphi_{R}^{2}(z)+\varphi_{L}^{2}(z))\\
&+\frac{1}{4}\eta_2\int dz\partial_zV(Q,z)(\varphi_{L}^{2}(z)-\varphi_{R}^{2}(z)).
\end{split}
\end{equation}
The Pauli form factor is given by
\begin{equation}
	\label{eq14}
F_{2}(Q^{2})=\eta_{2}\int dze^{-A(z)}V(Q,z)\varphi_{R}(z)\varphi_{L}(z),
\end{equation}
where $V(Q,z)$ represents the five-dimensional electromagnetic field.

The electromagnetic field action takes the form
\begin{equation}
	\label{eq15}
S=-\frac{1}{4}\int d^{4}xdz\sqrt{-g}e^{-\Phi(z)}F^{mn}F_{mn},
\end{equation}
where $F_{mn}=\partial_{m}V_{n}-\partial_{n}V_{m}$. The corresponding equation of motion may then be derived
\begin{equation}
	\label{eq16}
\partial_{m}(\sqrt{-g}e^{-\Phi(z)}F^{mn})=0.
\end{equation}
We impose the gauge condition
\begin{equation}
	\label{eq17}
\partial_{\mu}V^{\mu}+e^{-A(z)+\Phi(z)}\partial_{z}(e^{A_(z)-\Phi(z)}V_{z})=0.
\end{equation}
The equation of motion consequently takes the form
\begin{equation}
	\label{eq18}
\partial_{\mu}\partial^{\mu}V_{\nu}+(A'(z)-\Phi'(z))\partial_{z}V_{\nu}+\partial_{z}^{2}V_{\nu}=0,
\end{equation}
where $V_{\nu}=\epsilon_\nu e^{-iqx}V(Q,z)$. At the boundary, we impose $V(Q,0) = 1$ and implement the Neumann boundary condition in the infrared (IR) limit.

The electric and magnetic form factors can be expressed as
\begin{equation}
	\label{eq19}
G_E(Q^2)=F_1(Q^2)-\frac{Q^2}{4M^2}F_2(Q^2),
\end{equation}
\begin{equation}
	\label{eq20}
G_M(Q^2)=F_1(Q^2)+F_2(Q^2).
\end{equation}
\begin{figure}
	\centering
	\includegraphics[width=7cm]{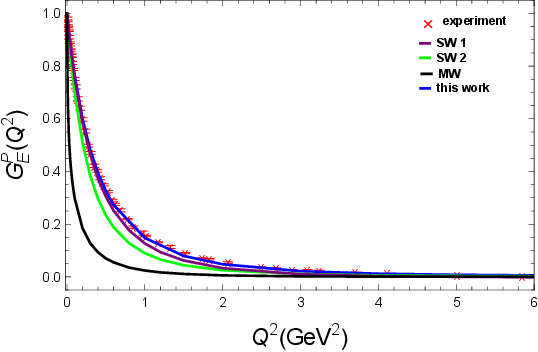}
	\caption{\label{figure}The proton electric form factor $G_E^P(Q^2)$. The red crosses represent experimental data from Ref. \cite{Arrington:2007ux}, the purple solid line and green solid line represent two soft-wall models \cite{Mamo:2021jhj, Mamo:2021krl}, the black solid line represents the modified warp factor model \cite{FolcoCapossoli:2019imm}, and the blue solid line represents our result.}
\end{figure}
\begin{figure}
	\centering
	\includegraphics[width=7cm]{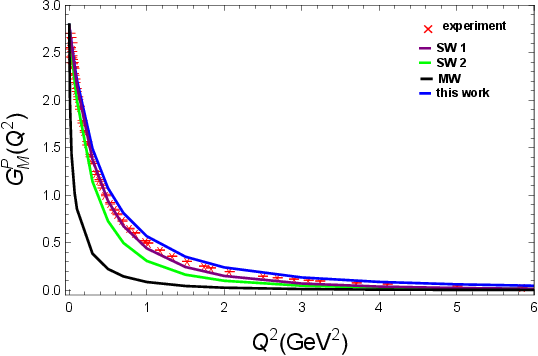}
	\caption{\label{figure}The proton magnetic form factor $G_M^p(Q^2)$. The red crosses represent experimental data from Ref. \cite{Arrington:2007ux}, the purple solid line and green solid line represent two soft-wall models \cite{Mamo:2021jhj, Mamo:2021krl}, the black solid line represents the modified warp factor model \cite{FolcoCapossoli:2019imm}, and the blue solid line represents our result.}
\end{figure}

Figure 2 and Figure 3 display the dependence of the electric and magnetic form factors on the momentum transfer $Q^2$, respectively. For the electric form factor, the average relative deviations from the experimental data for SW1, SW2, MW, and our model are 22.8\%, 28.5\%, 72.8\%, and 8.1\%, respectively; for the magnetic form factor, the average relative deviations for SW1, SW2, MW, and our model are 13.9\%, 28.6\%, 73.8\%, and 12.0\%, respectively. This indicates that our model captures the dominant features of the proton's electromagnetic structure. While these results are more precise than those obtained from the standard soft-wall model and the modified warp factor model, the remaining deviations suggest that richer physics and higher-order effects should be taken into account when studying the electromagnetic structure of the proton.

The notion of nucleon radii is fundamental to characterizing the internal architecture of protons and neutrons, offering essential insight into the nonperturbative domain of QCD. These radii are defined through multiple physical observables, each probing a distinct aspect of the nucleon’s internal organization. Beyond their importance in elucidating nucleon structure, they also play a vital role in precision tests of the Standard Model. In particular, high-precision measurements of nucleon radii provide stringent constraints on theoretical approaches such as lattice QCD, chiral effective field theory, and holographic QCD.

To support our theoretical treatment, the Breit frame is adopted, with the kinematical variables $\Delta^{\nu}$, $P^{\nu}$, and the momentum transfer squared defined as follows \cite{Polyakov:2018zvc}:
\begin{equation}
	\label{eq31}
P^{\nu}=(E,\textbf{0}), \ \Delta^{\nu}=(0,\bm{\Delta}),\  \Delta^2=4(M^2-E^2).
\end{equation}

The charge and magnetic radii represent key observables for probing the internal structure of nucleons. The charge radius quantifies the spatial extent of the electric charge distribution, whereas the magnetic radius characterizes the size associated with the magnetic moment, generated by the internal dynamics of quarks and gluons. The charge radius of the proton is defined as
\begin{equation}
	\label{eq32}
\langle r^2_E\rangle=\frac{\int d^3\textbf{r}r^2G_E(r)}{\int d^3\textbf{r}G_E(r)}=-6\frac{dG_E(Q^2)}{dQ^2}|_{Q^2=0}.
\end{equation}
The magnetic radius of the proton is defined as
\begin{equation}
	\label{eq33}
\langle r^2_M\rangle=\frac{\int d^3\textbf{r}r^2G_M(r)}{\int d^3\textbf{r}G_M(r)}=-\frac{6}{\mu}\frac{dG_M(Q^2)}{dQ^2}|_{Q^2=0},
\end{equation}
where $\mu$ represents the magnetic moment of the proton. With $r_E^p = 0.812 fm$ and $r_M^p = 0.762 fm$, this result is relatively close to the lattice QCD result but shows a slight deviation from the experimental measurement, as shown in Figure 4.
\begin{figure}
	\centering
	\includegraphics[width=8.5cm]{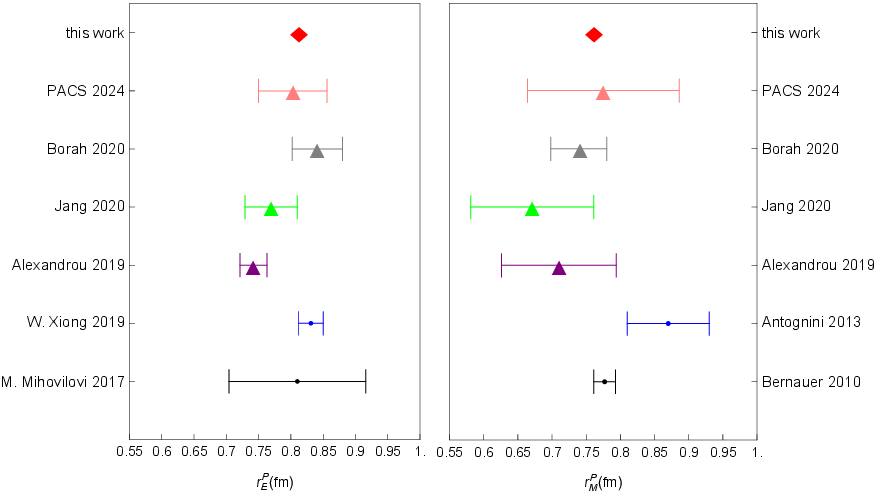}
	\caption{\label{figure}A comparison of the proton's charge and magnetic radii with other models \cite{Tsuji:2023llh, Borah:2020gte, Jang:2019jkn, Alexandrou:2018sjm} and experiments \cite{Mihovilovic:2016rkr, Xiong:2019umf, A1:2010nsl, Antognini:2013txn}: the left side shows the charge radius, and the right side displays the magnetic radius.}
\end{figure}

\section{Gravitational form factors}\label{sec:04}

The generation of the proton's mass represents one of the most fundamental challenges in QCD. Remarkably, the QCD Lagrangian contains only light quarks and massless gluons as fundamental degrees of freedom, yet must generate the observed proton mass of approximately 0.94 GeV. A corresponding fundamental question addresses the origin of the proton's spin: although simplistic quark models attribute its spin-1/2 configuration exclusively to the intrinsic spins of valence quarks, the complete theoretical decomposition of the proton's total angular momentum continues to present an outstanding challenge in hadronic physics. To address these fundamental questions, we examine the matrix elements of the proton's energy-momentum tensor (EMT). These matrix elements provide direct access to the proton's core structural characteristics, including its mass distribution, spin composition, and internal stress profiles \cite{Burkert:2023wzr}. 

The general parameterization of the EMT matrix elements for the proton can be written as
\begin{equation}
\begin{split}
	\label{eq19}
\langle p'|T^{\mu\nu}(0)|p\rangle=&\bar{u}(p')[\gamma^{(\mu}P^{\nu)}A(Q^{2})+\frac{iP^{(\mu}\sigma^{\nu)\alpha}k_{\alpha}}{2M}B(Q^{2})\\
&+\frac{k^{\mu}k^{\nu}-\eta^{\mu\nu}k^{2}}{4M}D(Q^{2})]u(p),
\end{split}
\end{equation}
where $\gamma^{(\mu}p^{\nu)}=\frac{1}{2}(\gamma^{\mu}p^{\nu}+\gamma^{\nu}p^{\mu})$, $k^{2}=(p'^\mu-p^\mu)^2$, and $P^\mu=\frac{1}{2}(p^\mu+p'^\mu)$. In holographic QCD, Equation (24) serves as the source term for metric fluctuations, $\eta'_{\mu\nu}=\eta_{\mu\nu}+h_{\mu\nu}$. The gravitational form factors of the proton are obtained through the coupling between the irreducible representations of the five-dimensional graviton field $h_{\mu\nu}$ and the bulk proton field. The energy-momentum tensor of the proton can be expressed as
\begin{equation}
	\label{eq20}
T^{\mu\nu}(x,z)=\frac{-2}{\sqrt{-g}}\frac{\delta \mathcal{L}}{\delta g_{\mu\nu}}.
\end{equation}
Substituting the metric perturbation $\eta'_{\mu\nu}=\eta_{\mu\nu}+h_{\mu\nu}$ into the action and performing a perturbative expansion yields the interaction action between gravitons and protons
\begin{equation}
	\label{eq21}
S_{int}=\frac{1}{2}\int d^4xdz\sqrt{-g}h_{\mu\nu}T^{\mu\nu}+ \mathrm{o}(h^2).
\end{equation}
Substituting the perturbed metric $\eta'_{\mu\nu}= \eta_{\mu\nu}+h_{\mu\nu}$ into the gravitational action $S_G$, we obtain the corresponding action as
\begin{equation}
\begin{split}
	\label{eq22}
S_h=\frac{1}{4\kappa^2}\int d^4xdz\sqrt{-g}e^{-2\Phi(z)}(\partial_\alpha h^{\mu\nu}\partial^\alpha h_{\mu\nu}-\frac{1}{2}\partial_\alpha h\partial^\alpha h). 
\end{split}
\end{equation}
Introduce the harmonic-traceless gauge $\partial_\lambda h^{\lambda}_\alpha=\partial_\alpha h=0$. Under this gauge condition, with $h=h^\mu_\mu$ denoting the trace and $\kappa$ as the Newton constant, the graviton satisfies the equation of motion given by Eq. 27:
\begin{equation}
	\label{eq23}
\partial_{\alpha}[\sqrt{-g}e^{-2\Phi(z)}g^{\alpha\lambda}\partial_{\lambda}h_{\mu\nu}]=0.
\end{equation}
This can then be simplified to
\begin{equation}
	\label{eq23}
(\partial_{z}^{2}+(3A'(z)-2\Phi'(z))\partial_{z}-Q^2)h(Q^2,z)=0,
\end{equation}
With the ansatz $h_{\mu\nu}=\epsilon_{\mu\nu}e^{-ikx}h(Q^2,z), k^2=-Q^2$, we impose the following boundary conditions on $h(Q^2,z)$:
\begin{equation}
	\label{eq30}
h(Q^2,0)=h(0,z)=1,\ \ \partial_zh(Q^2,\infty)=0.
\end{equation}

From the above analysis, we can obtain
\begin{equation}
	\label{eq28}
A(Q^{2})=\frac{1}{2}\int dzh(Q^2,z)(\varphi_{L}^{2}(z)+\varphi_{R}^{2}(z)).
\end{equation}
\begin{figure}[H]
	\centering
	\includegraphics[width=7cm]{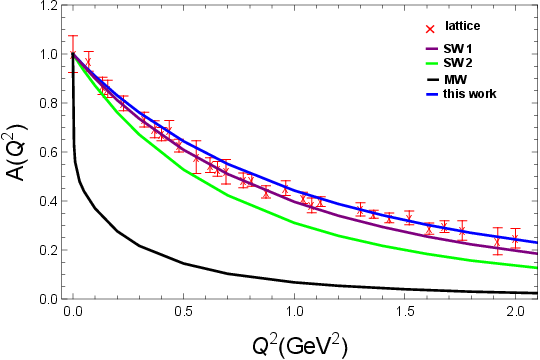}
	\caption{\label{figure}The proton gravitational form factor $A(Q^2)$. The red crosses represent lattice data at $\mu^2 = 4\ \mathrm{GeV}^2$ \cite{Hackett:2023rif}, the purple solid line and green solid line represent two soft-wall models \cite{Mamo:2021jhj, Mamo:2021krl}, the black solid line represents the modified warp factor model \cite{FolcoCapossoli:2019imm}, and the blue solid line represents our result.}
\end{figure}

As shown in Figure 5, we compared our results with those from other models and with lattice calculations. Our results are also closer to the lattice calculations. The average relative deviations from the lattice data for SW1, SW2, MW, and our model are 6.4\%, 22.6\%, 77.1\%, and 3.9\%, respectively.

In Ref. \cite{Mamo:2021krl}, the bulk metric fluctuation can be decomposed as 
\begin{equation}
\begin{split}
	\label{eq19}
h_{\mu\nu}(k,z)=&[\epsilon_{\mu\nu}^{TT}h(k,z)+k_\mu k_\nu H(k,z)]\\
&+[k_\mu A_\nu^\perp(k,z)+k_\nu A_\mu^\perp(k,z)]+\frac{1}{3}\eta_{\mu\nu}f(k,z),
\end{split}
\end{equation}
where $h$ and $H$ correspond to the spin-2 transverse traceless and longitudinal trace parts, respectively, $A_\nu^\perp$ denotes the spin-1 transverse vector part, and $f$ denotes the spin-0 trace part.
Using the spin-2 transverse traceless polarization tensor contracted with the energy-momentum tensor, one obtains
\begin{equation}
	\label{eq30}
\langle p_2|\epsilon_{\mu\nu}^{TT}T^{\mu\nu}(0)|p_1\rangle=\bar{u}(p_2)(A(Q^2)\epsilon_{\mu\nu}^{TT}\gamma^\mu p^\nu)u(p_1),
\end{equation}
with
\begin{equation}
	\label{eq30}
A(Q^2)=\frac{1}{2}\int dz(\varphi_R^2(z)+\varphi_L^2(z))h(Q^2,z).
\end{equation}
Using the trace part contracted with the energy-momentum tensor, one obtains
\begin{equation}
\begin{split}
	\label{eq19}
\frac{1}{3}\langle p_2|\eta_{\mu\nu}T^{\mu\nu}(0)|p_1\rangle &=\bar{u}(p_2)(A(Q^2)\frac{m_N}{3}+\frac{Q^2}{m_N}C(Q^2))u(p_1)\\
&=\bar{u}(p_2)(A_S(Q^2))u(p_1),
\end{split}
\end{equation}
with
\begin{equation}
	\label{eq30}
A_S(Q^2)=\frac{\tilde{C}}{2}\int dz(\psi_R^2(z)+\psi_L^2(z))f(Q^2,z).
\end{equation}
In our model, $B(Q^2)=0$ holds as a general property, and this result is further confirmed by the vanishing of the coupling due to the spin connection. 
$C(Q^2)=\frac{1}{4}D(Q^2)$ can be written as
\begin{equation}
	\label{eq30}
C(Q^2)=\frac{1}{3}\frac{m_N^2}{Q^2}(A(Q^2)-A_S(Q^2)).
\end{equation}
The forms of $A(Q^2)$ and $A_S(Q^2)$ are assumed to be\cite{Mamo:2021krl},
\begin{equation}
	\label{eq30}
A(Q^2)=\frac{1}{(1+\frac{Q^2}{m_T^2})^2}, \ \ A_S(Q^2)=\frac{1}{(1+\frac{Q^2}{m_S^2})^2}.
\end{equation}
The parameters are determined as \( m_T \) from the gravitational form factor \( A \), and \( m_S\) from the lattice gravitational form factor \( D(0) \).
\begin{figure}[H]
	\centering
	\includegraphics[width=8cm]{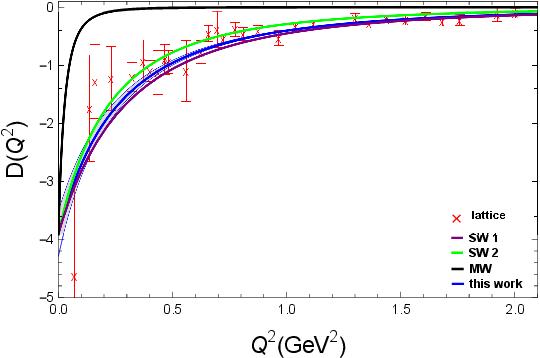}
	\caption{\label{figure}The proton gravitational form factor $D(Q^2)$. The red crosses represent lattice data at $\mu^2 = 4\ \mathrm{GeV}^2$\cite{Hackett:2023rif}, the purple solid line and green solid line represent two soft-wall models \cite{Mamo:2021jhj, Mamo:2021krl}, the black solid line represents the modified warp factor model \cite{FolcoCapossoli:2019imm}, and the blue solid line represents our result.}
\end{figure}

We have computed the proton gravitational form factor $D(Q^2)$ by employing the holographic method of Ref. [26] together with our modified dilaton field. We fit our gravitational form factor $A(Q^2)$ to obtain $m_T \approx 1409$~MeV, and fix $m_S \approx 679$~MeV from the lattice value $D(0) \approx 3.9$. As shown in Figure~6, our results are in reasonable agreement with the lattice data, where the shaded area indicates the uncertainty corresponding to a $\pm 10\%$ variation in $D(0)$. We compared our results with those from other models and with lattice calculations. The average relative deviations from the lattice data for SW1, SW2, MW, and our model are 34.1\%, 25.2\%, 97.8\%, and 25.9\%, respectively. According to the average relative deviation, the agreement between our result and the lattice data is comparable to that between the SW2 model and the lattice data, and better than that of other models.

In our model, the extracted glueball masses are $m_T\approx1409$ MeV and $m_S \approx 679$ MeV. The distinct values of these masses imply a non-degenerate glueball spectrum, which, following the holographic mechanism of Ref.[26], generates a non-trivial D-term. The resulting non-degeneracy is a physical feature of our model, consistent with the holographic mechanism of Ref.[26].

The energy, pressure, and shear force distributions within the proton are expressed in terms of combinations of its gravitational form factors in the following \cite{Polyakov:2018zvc}:
\begin{equation}
	\label{eq34}
\epsilon(r)= M[A(Q^{2})+\frac{Q^{2}}{4M^2}(A(Q^{2})-2J(Q^{2})+D(Q^{2})]_{FT}.
\end{equation}
\begin{equation}
	\label{eq35}
p(r)= \frac{1}{6M}\frac{1}{r^2}\frac{d}{dr}(r^2\frac{d}{dr}[D(Q^{2})]_{FT}).
\end{equation}
\begin{equation}
	\label{eq36}
s(r)= -\frac{1}{4M}r\frac{d}{dr}(\frac{1}{r}\frac{d}{dr}[D(Q^{2})]_{FT}).
\end{equation}
where FT denotes the Fourier transform and takes the form:
\begin{equation}
	\label{eq37}
[f(r)]_{FT}=\int \frac{d^3\bm{\Delta}}{(2\pi)^3}e^{-i\bm{\Delta}\cdot \textbf{r}}f(Q^{2}).
\end{equation}

The mass radius of the proton is
\begin{equation}
	\label{eq39}
\langle r^2_{mass}\rangle=\frac{\int d^3\textbf{r}r^2\epsilon(r)}{\int d^3\textbf{r}\epsilon(r)}=-6\frac{dA(Q^{2})}{dQ^{2}}|_{Q^{2}=0}-\frac{3D(0)}{2M^2}.
\end{equation}
The longitudinal force is given in terms of $p$ and $s$ as:
\begin{equation}
	\label{eq40}
F_{||}(r)=p(r)+\frac{2}{3}s(r),
\end{equation}
where it represents the normal force per unit area. Local stability requires the longitudinal force to be positive ($F_{||}(r)>0$), indicating an outward direction; otherwise, the proton would undergo collapse. Similarly, in stable hydrostatic systems, the shear force is generally positive, satisfying $s(r)>0$ and thereby providing a second condition for local stability. The mechanical radius of the proton is
\begin{equation}
	\label{eq41}
\langle r^2_{mech}\rangle=\frac{\int d^3\textbf{r}r^2F_{||}(r)}{\int d^3\textbf{r}F_{||}(r)}=\frac{6D(0)}{\int^0_\infty D(t)dt}.
\end{equation}

Our extracted values for the proton mass radius \( r_{\text{mass}} = 0.713(19) \, \text{fm} \) and mechanical radius \( r_{\text{mech}} = 0.741(17) \, \text{fm} \) are in agreement with lattice QCD determinations and other theoretical model predictions. The uncertainties are obtained from a stability analysis by varying \(D(0)\) by \(\pm 10\%\).

By substituting the obtained gravitational form factors into Equations (39)–(41) and (44), we obtain the energy, pressure, shear, and longitudinal force distributions of the proton, as presented in Figs. 7.

Our calculations satisfy the energy conservation and mechanical equilibrium conditions:
\begin{equation}
	\label{eq39}
M=\int d^3\bm{r}\epsilon(r),\ \  \ \int dr r^2p(r)=0. 
\end{equation}

\begin{figure}
	\centering
	\includegraphics[width=8.5cm]{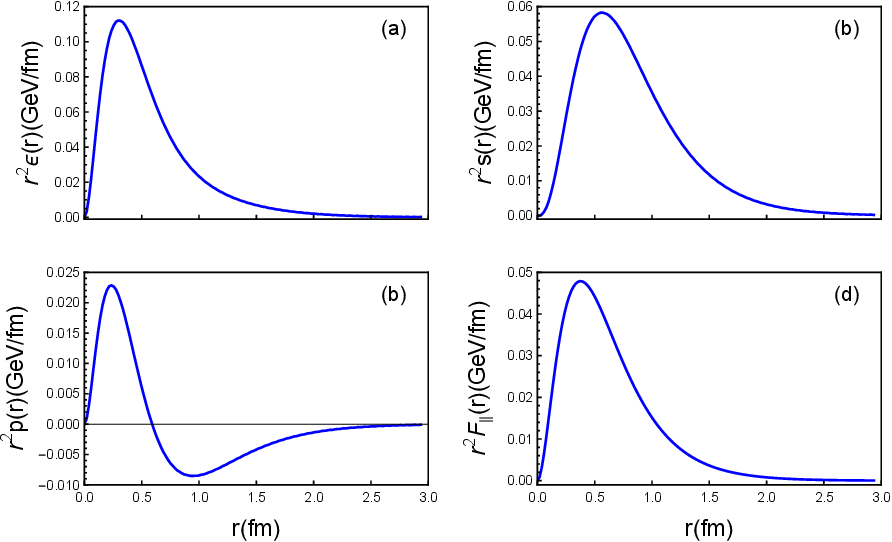}
	\caption{\label{figure} Spatial distributions of the weighted (a) energy density $r^2\epsilon(r)$, (b) pressure $r^2p(r)$, (c) shear force $r^2s(r)$, and (d) longitudinal force $r^2F_{\parallel}(r)$ inside the proton.}
\end{figure}

As shown in Fig.~7, panel (a) represents the weighted energy density, which has a peak at $r \approx 0.30$~fm, and its integral satisfies the energy conservation (Eq.~46). Panel (b) represents the weighted pressure. Unlike the weighted energy density, which remains positive throughout, the weighted pressure distribution exhibits a sign change at approximately $r = 0.59$~fm. The positive pressure peaks at $r = 0.23$~fm, while the negative pressure reaches its maximum magnitude at about $r = 0.92$~fm. The pressure adheres to the von Laue condition in Eq. (46), where the outward push from the positive region is balanced by the inward pull from the negative region, thereby ensuring the proton’s mechanical stability. The positive pressure occupies a compact core, whereas the negative pressure spreads out over an extended periphery. Panel (c) represents the weighted shear force, which has a peak at $r \approx 0.56$~fm and reflects the internal stress anisotropy of the proton. Panel (d) represents the weighted longitudinal force , which has a peak at $r \approx 0.38$~fm and is directly related to the local force balance. The relationship between the longitudinal force, the pressure, and the shear force is discussed in Eq.~44.

In this section, we first predict the proton's gravitational form factor $A(Q^2)$ from our model, and the results are consistent with lattice calculations. In our model, $B(Q^2)=0$ holds as a general property, and this result is further confirmed by the vanishing of the coupling due to the spin connection, which is also consistent with lattice calculations indicating that this term is very small. Subsequently, we decompose the graviton into a transverse traceless part and a trace part, and obtain the expression for the gravitational form factor $D(Q^2)$. The parameter is determined from the lattice value of $D(0)$. The resulting $D(Q^2)$ is reasonably consistent with the lattice data. Finally, using the gravitational form factors as inputs, we investigate four different mechanical quantities inside the proton.

To facilitate a direct comparison among the different soft-wall variants across all observables, we summarize in Table~II the average relative deviations from the experimental and lattice data for the proton mass spectrum, electromagnetic form factors, and gravitational form factors obtained with SW1, SW2, MW, and our model.
\begin{table}[H]
	\centering
	\includegraphics[width=8.5cm]{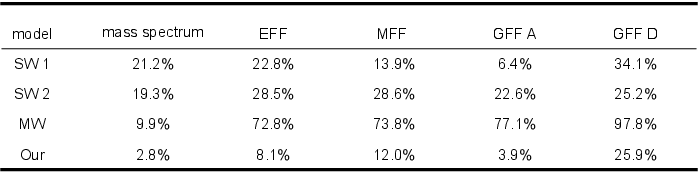}
	\caption{\label{Table II}Relative deviations from experimental and lattice data for different holographic models across all observables.}
\end{table}
The comparison shows that our model achieves the smallest or comparable relative deviations in all individual observables, and provides the best overall average performance among the soft-wall variants considered.

\section{Conclusion}\label{sec:07}

In the study of proton structure, we employ a soft-wall model that incorporates a dilaton field $\Phi(z)=k_1^2z^2tanh(k_2^4z^2/k_1^2)$. This model breaks conformal symmetry in the low-energy region to generate confinement effects and simulates gluon condensation in the high-energy region, thereby more closely approximating the actual behavior of quantum chromodynamics compared to the traditional $\Phi(z)=k_3^2z^2$-type soft-wall model. In this work, we investigate the proton structure by calculating the mass spectrum, electromagnetic form factors, electromagnetic radius, and gravitational form factors. Using these gravitational form factors as input, we then obtain the mechanical properties of the proton.

We begin by calculating the proton mass spectrum using a Schrödinger-like equation incorporating anomalous dimensions. The results exhibit a deviation of less than 3\% from experimental data, demonstrating good consistency.

Subsequently, using the same parameters, we predict the electromagnetic form factors, electromagnetic radius, and gravitational form factor $A(Q^2)$ of the proton. These predictions are found to be consistent with experimental measurements and lattice calculations. 

Next, we separate the graviton into a transverse traceless part and a trace part, which allows us to obtain an expression for $D(Q^2)$. The only parameter is determined from the lattice input $D(0)$. Our calculated $D(Q^2)$ is found to be reasonably consistent with the lattice data. Finally, with the gravitational form factors in hand, we evaluate the spatial distribution of mechanical properties within the proton.

Furthermore, we have compared our results with those of the soft-wall model and the modified warp factor model. Our model exhibits the smallest relative average deviation, demonstrating that it can simultaneously describe these observables with higher accuracy than the these models.

Although the results are encouraging, it is essential to acknowledge several limitations of this study. For instance, some discrepancies remain between our calculated magnetic form factor and electromagnetic radius and the corresponding experimental measurements. Future research should aim to clarify these discrepancies and explore the long-term implications of our findings. Furthermore, applying this framework to study other hadrons and their interactions may contribute to a deeper understanding within the broader context of quantum chromodynamics.

In summary, the results obtained from our model are generally consistent with both lattice calculations and experimental measurements. Our results show that a relatively simple phenomenological holographic model—with its parameters determined by the proton mass spectrum—is able to predict the electromagnetic form factors, the electromagnetic radius, and the gravitational form factor $A(Q^2)$ that are in basic agreement with experimental data and lattice results. These findings offer some useful perspectives for future theoretical studies of the proton's internal structure.

\section*{Acknowledgments}

We thank Hai-cang Ren and  Yan-Qing Zhao for useful discussions. This work is supported in part by the National Key Research and Development Program of China under Contract No. 2022YFA1604900. This work is also partly supported by the National Natural Science Foundation of China(NSFC) under Grants No. 12435009, and No. 12275104.

\section*{References}

\bibliography{ref}
\end{document}